\documentclass[prb,aps,twocolumn,amsmath,amssymb,showpacs]{revtex4}

\usepackage{graphicx}
\usepackage{dcolumn}
\usepackage{bm}

\begin{document}
\title{Influence of doping level on the Hall coefficient and on the
    thermoelectric power in $Nd_{2-x}Ce_xCuO_4$}

\author{C. H. Wang, G. Y. Wang, T. Wu, Z. Feng, X. G. Luo}
\author{X. H. Chen}
\altaffiliation{Corresponding author} \email{chenxh@ustc.edu.cn}
\affiliation{Hefei National Laboratory for Physical Science at
Microscale and Department of Physics, University of Science and
Technology of China, Hefei, Anhui 230026, People's Republic of
China\\ }

\date{\today}

\begin{abstract}
Hall coefficient $R_H$ and thermoelectric power TEP are studied
systematically in the single crystals $Nd_{2-x}Ce_xCuO_4$ (NCCO)
with different x from underdoped to overdoped regime. $R_H$ and
TEP decrease and change their sign from negative to positive with
increasing doping level x. A striking feature is that the
temperature dependence of the Hall angle follows a $T^4$ behavior
in the underdoped regime, while a $T^2$ law in the overdoped
regime. This behavior is closely related to the evolution of Fermi
surface with doping level observed by the angle-resolved
photoemission spectroscopy (ARPES).

\end{abstract}

\vskip 15 pt


 \narrowtext

\maketitle

$Nd_{2-x}Ce_{x}CuO_{4+\delta}$ (NCCO) belongs to a quite
interesting class of materials which is called as electron doped
cuprates. With the substitution of Nd by Ce,  the electrons were
injected to the $CuO_2$ plane since the Hall coefficient ($R_H$)
and the thermoelectric power (TEP) remains negative.\cite{Onose}
However, $R_H$ and thermoelectric power gradually increase and
eventually change sign to positive with further doping
Ce.\cite{Hagen,Seng,JiangR,JiangS,Wangzz,Yanghs} Similar feature
has been observed with changing the oxygen
content.\cite{JiangR,Fournier} In another n-type cuprate
$Pr_{2-x}Ce_{x}CuO_{4+\delta}$ (PCCO), the change of $R_H$ sign
with temperature\cite{Brinkmann01,Fournier98,Brinkmann02} and
doping\cite{Dagan} were also observed.  Those strange behaviors
strongly indicate that at moderate doping, NCCO and PCCO have two
bands with different types of charge carriers, one is hole carrier
and the other is electron carrier. The Fermi surface (FS) obtained
by ARPES indirectly supports this conclusion. Armitage et al.
\cite{Armitage} suggested that the existence of FS patch in the
slightly doped samples is marked by an electron pocket at
$(\pi,0)$ and the FS patch gradually changes to a large hole-like
FS with the appearance of section near the zone diagonal
$(\pi/2,\pi/2)$ of the Brillouin zone with increasing doping.  The
presence of the two separate FS pockets may result from the band
folding effect induced by the antiferromagnetic
correlations.\cite{yuan}  The theoretical calculations indicate
that the two FS pockets can be effectively described as a two-band
system.\cite{yuan,kusko} Recently,  Luo and Xiang proposed a
weakly coupled two-band model with $d_{x^2-y^2}$ pairing symmetry
to account for the anomalous temperature dependence of superfluid
density ($\rho_s$) in n-type cuprate superconductor very
well.\cite{luo}

Hall angle is another interesting feature. It is well known that
Hall angle follows a $T^2$ law in hole doped system.
Anderson\cite{Anderson01} emphasized that the transport is
governed by two different scattering time for the high $T_c$
cuprates, $\tau_{tr}(T^{-1})$ for in-plane resistivity and
$\tau_H(T^{-2})$ for Hall angle. Another point of view is that the
cotangent of Hall angle ($cot\theta_H$) is proportional to the
square of the scattering rate which can be independently measured
by the zero-field resistivity.\cite{Varma} Recently, a striking
finding by Ando et al.\cite{Ando01} is that the temperature
dependence of resistivity nearly follows $T^2$ in lightly doped
YBCO and LSCO. Their finding gives another picture that in lightly
hole-doped cuprates the transport behavior is Fermi-liquid-like
and results from the quasiparticles on the Fermi arcs. As we know,
the resistivity is also not T-linear in electron doped cuprates.
In comparison with hole doped cuprates, it is quite interesting to
study the Hall angle and its relation to the resistivity in the
NCCO system. Although the nearly $T^4$ behavior of Hall angle has
been mentioned by Fournier et al.\cite{Fournier} in optimum doped
$Nd_{1.85}Ce_{0.15}CuO_{4-\delta}$ thin film, Hall angle behavior
was still not clear in n-type cuprates yet. Therefore, it is quite
meaningful to study variation of Hall angle behavior with
temperature and doping in this kind of system.

In this report, the Hall effect and thermoelectric power were
systematically studied. The Hall contact configuration is standard
ac six-probe geometry with magnetic field up to 10 Tesla. The Hall
signal was extracted from the antisymmetric part of the transverse
signal measured with the opposite field direction to remove the
longitudinal contribution due to the misalignment of the Hall
voltage contact point. The TEP measurements were performed using
small and reversible temperature difference of 0.2-0.5 K. Two ends
of the single crystals were attached to two separated copper heat
sink to generate the temperature gradient along the crystal
ab-plane. Two Rh-Fe thermometers were glued to the heat sink (just
next to the single crystals). Copper leads were adhered to the
single crystals and all the data were corrected for the
contribution of the Cu leads.

As shown in figure 1(a), the resistivity shows an upturn at the
low temperatures for the underdoped samples. Especially for
x=0.025, such low temperature insulating behavior becomes obvious
around 240 K, similar to the report by Onose et al.\cite{Onose}
However, the metallic behavior can be observed in the high
temperature range. For the overdoped samples, metallic behavior
exists in the whole temperature range and superconductivity was
observed for the crystal with x=0.17. The temperature dependent
Hall coefficient and Hall mobility ($\mu_H=R_H/\rho_{ab})$ are
plotted in fig.1(b) and (c), respectively.

\begin{figure}[t]
\centering
\includegraphics[width=9cm]{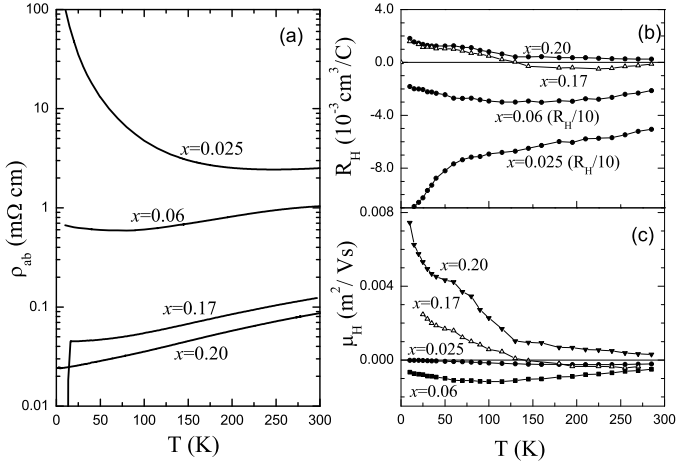}
\caption{\label{fig:1} Temperature dependence of (a): in-plane
 resistivity ($\rho_{ab}$); (b): Hall coefficient $R_H$;
and (c): Hall mobility  ($\mu_H=R_H/\rho_{ab})$.}
\end{figure}

Hall coefficient and Hall mobility show a continuous variation
with temperature. $R_H$ and $\mu_H$ are negative in the whole
temperature range for nonsuperconducting, underdoped samples with
x=0.025 and 0.06, while positive for nonsuperconducting, overdoped
sample with x=0.20. For x$\sim$0.17, $R_H$ and $\mu_H$ change
their sign at a certain temperature with decreasing temperature.
Such doping dependent behavior is similar to that observed in
$Nd_{1.85}Ce_{0.15}CuO_{4+\delta}$ films \cite{JiangR,Fournier}
and $Pr_{1.85}Ce_{0.15}CuO_{4+\delta}$ single
crystals\cite{Brinkmann01,Brinkmann02} by varying oxygen content.
Furthermore, in optimum doped (x$\sim$0.15) and slightly overdoped
superconducting (x$\sim$0.17) compositions, the change of Hall
coefficient sign with temperature has been
reported\cite{Wangzz,JiangR,Fournier,Fournier98}. The conventional
single-carrier transport can not explain such behavior adequately,
and two types of the carriers have to be used to explain these
behaviors. $R_H$ shows a downturn for x=0.025, but for x=0.06 the
absolute value of $R_H$ slightly decreases. This case is similar
to the behavior in lightly doped LSCO system, in which $R_H$
increases in low temperatures for low doping level, while slightly
decreases when doping level up to 0.07 \cite{Ando01}.

\begin{figure}[t]
\centering
\includegraphics[width=9cm]{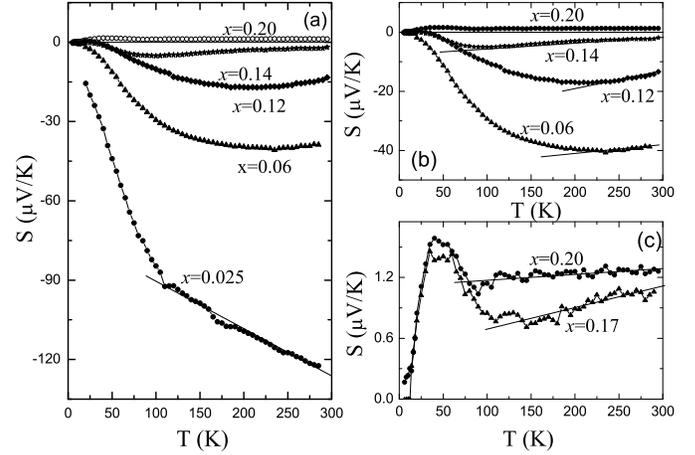}
\caption{\label{fig:2} (a)£ºTemperature dependence of
thermoelectric power (TEP) for the $Nd_{2-x}Ce_xCuO_4$ crystals
with different doping level; (b): TEP vs. T for x=0.06, 0.12,
0.14, 0.20; (c): TEP vs. T for the overdoped samples.}
\end{figure}

Figure 2(a) shows the temperature dependent thermoelectric power
(TEP) for underdoped and overdoped samples. TEP monotonously
decreases with increasing doping level. For the overdoped sample
the TEP is positive. The TEP has a similar behavior to the doping
dependent tendency of $R_H$. It further indicates that two types
of the carriers exist in the system. For x=0.025 sample, the TEP
monotonically increases with increasing temperature and no
saturation is observed. When the doping up to 0.06, a broad peak
of TEP is observed. Similar behavior also occurs in the crystals
with x=0.12 and 0.14 as shown in Fig.2(b). The temperature
corresponding to the peak decreases with increasing the doping
level. The similar behavior has been reported in
$Sm_{2-x}Ce_xCuO_{4+\delta}$ (SCCO) polycrystalline
samples\cite{Yanghs}, NCCO single crystals,\cite{Xuxq} and
$Nd_{1.85}Ce_{0.15}CuO_{4+\delta}$ thin films with changing the
oxygen content\cite{Fournier}. Fig.2(c) shows temperature
dependent TEP for the crystals with x=0.17 and 0.20. The small
magnitude of TEP is typical metal.

In the polycrystaline $Sm_{2-x}Ce_xCuO_{4+\delta}$
 (SCCO)\cite{Yanghs}, the TEP sign changes with decreasing
temperature for the overdoped SCCO samples, while TEP remains
positive in our overdoped crystal, and no sign change was
observed. This case is similar to the optimum sample that no sign
change is observed in NCCO thin film\cite{Fournier}, while sign
change is observed in PCCO polycystalline sample\cite{Leesi}. The
sign change for polycystalline samples may result from the
inhomogeneity of the oxygen content in the grains.

\begin{figure}[t]
\centering
\includegraphics[width=9cm]{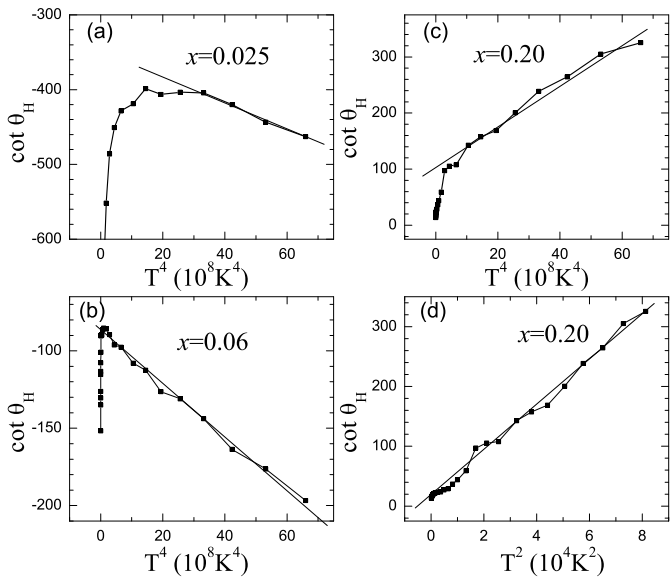}
\caption{\label{fig:4} Temperature dependence of $cot\theta_H(\sim
 \rho_{ab}/R_H$) at 10 T in $T^4$ scale for the crystals $Nd_{2-x}Ce_xCuO_4$(a): x=0.025, (b): x=0.06, (c):
 x=0.20, (d): x=0.20 in $T^2$ scale.}
\end{figure}

In fig.3, the cotangent of Hall angle as a function of temperature
is plotted for the crystals with x=0.025, 0.06 and 0.20. In
fig.3(a) and (b), a $T^4$ power law behavior is clearly observed
above 240 K and 90 K for the samples with x=0.025 and 0.06,
respectively. The downturn around 240 K and 90 K coincides with
the low-temperature upturn in the resistivity $\rho_{ab}$ shown in
fig.1(a). The $T^4$ law behavior in the crystals with x=0.025 and
x=0.06 can not be explained by Anderson's two dimensional
Luttinger liquid theory as it predicts. the $T^2$ dependent
cotangent of Hall angle based on spinon-spinon
scattering\cite{Anderson01}. The resistivity in NCCO system has
closely a $T^2$ dependence except for the upturn of the low
temperature $\rho$ until the pseudogap opening
temperature\cite{Onose,Chenxh} for the sample x=0.06. Varma and
Abrahams's theory seems to be reasonable for $T^4$ law since they
predicted that the $cot\theta_H$ should follow the square of the
resistivity dependence\cite{Varma}. Wood et al. have suggested
this kind of possibility because they found that $cot\theta_H$
followed $T^{3.4}$ behavior with $\rho_{ab} \propto T^{1.7}$ in
their ion-irradiated NCCO thin films\cite{Wood}. Following the
argument of Ando et al. in $LSCO$ system, we can come to another
picture that the $T^4$ law might result from electron pocket (a
small FS around $(\pi,0)$) as the doped electrons concentrate to
this electron pocket in lightly doped NCCO. As we will see below,
the electron pocket governs the transport behavior in lightly
doped NCCO, which is the same way as that the Fermi arc dominates
the transport behavior in lightly doped
$LSCO$\cite{Ando01,Yoshida}. Therefore, the electron pocket must
play an important role to explain this strange $T^4$ law. The
difference between $T^4$ law in NCCO and $T^2$ law in hole-doped
cuprates suggests that the property of the electron pocket is
quite different from Fermi arc, and confirms the particle-hole
asymmetry in some sense. However, the data deviate from the $T^4$
law for x=0.20 sample as shown in Fig.3(c), a T-square dependent
the Hall angle is observed in whole temperature range shown in
fig.3(d). This observation agrees to the above picture, that is:
the electron pocket has already deformed and a large hole-like FS
begins to emerge for the overdoped sample. Therefore, the large
hole-like pocket at ($\pi/2$,$\pi/2$) dominates the charge
transport, so that a $T^2$ law observed in hole-doped cuprates
appears in the heavily doped n-type cuprates. If the quartic law
is a universal law for electron-type band, one can consider the
possibility to extract scattering rates for the two bands from the
data of Hall angle. Another interesting idea is that one may link
this change of Hall angle with the possible quantum critical point
on the analogy of their hole-doped partners
\cite{Orenstein,Ando01,Ando02}. However, one difficulty is that
the sign of Hall coefficient changes, and it is very hard to get
the temperature dependent Hall angle around $x=0.15 \sim 0.18$.

\begin{figure}[t]
\centering
\includegraphics[width=0.4\textwidth]{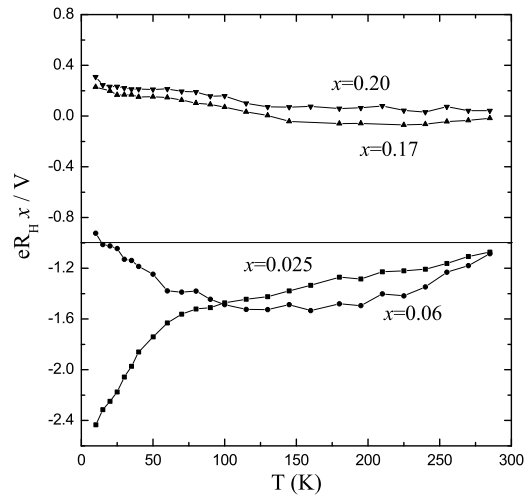}
\caption{\label{fig:4} $eR_Hx/V$ as a function of temperature for
the $Nd_{2-x}Ce_xCuO_4$ cystals with x=0.025, 0.06, 0.17 and
0.20.}
\end{figure}

It is believed that for the electron-doped cuprates two different
types of bands take effect in the transport behavior of those
samples as emphasized by many groups\cite{JiangR,Seng,Wangzz}.
Here, let us look at it from a different point of view. Fig. 4
shows the variation of $eR_Hx/V$ with T, where e is electron
charge and V is unit volume per Cu. The value should be -1 if the
nominal electron density x approximately accounts for the carrier
density. From the plot, it is easy to find that $eR_Hx/V$ is very
close to -1 around 300 K for the crystals with x=0.025 and 0.06.
The absolute value of $eR_Hx/V$ decreases significantly with
increasing x to the overdoping regime. The sign of $eR_Hx/V$
changes around 130 K for the sample with x=0.17, while remains
positive in the whole temperature range for the sample of x=0.20.
Since the largest volume of FS is proportional to 1+x for the
electron-doped samples, the effective carrier density should be
bounded by 1+x and the theoretical value of $eR_Hx/V$ should not
be smaller than $\frac{x}{1+x}$ in single-band model, so that the
theoretical lower limit is around 0.15 for x=0.17 and 0.17 for
x=0.20. But the actual value of $eR_Hx/V$ is approximately -0.01
for the sample with x=0.17 at room temperature and 0.04 for the
sample with x=0.20. Therefore, one can conclude that the single
band model is not valid in these cases. In the frame of two-bands
model\cite{JiangR,Seng,Wangzz}, $eR_Hx/V$ is proportion to
$\frac{n_p\mu_p^2-n_e\mu_e^2}{(n_p\mu_p+n_e\mu_e)^2}$ ( $n_p$,
$n_e$ are the carrier density of different bands. $\mu_p$, $\mu_e$
are the mobility of each band, respectively). According to the
above formula, our results can be well understood if we suppose
that the value of $n_p\mu_p^2$ is very close to $n_e\mu_e^2$ at
room temperature for our sample x=0.17, and increases with
increasing the doping level.

It is not clear why the two bands exist in the NCCO system.
However, ARPES results by Armitage et al.\cite{Armitage} help us
to understand this behavior. The FS patch at $(\pi/2,0)$was
observed in lightly doped NCCO and makes the $R_H$ negative. It
gradually deforms and the positive curvature part of the FS around
$(\pi,\pi)$ begins to increase with increasing Ce, and eventually
a large hole-like FS eventually appears. This part may give a
positive contribution to $R_H$. Therefore, the two bands model can
be phenomenally understood as a competition between electron-like
FS and hole-like FS. Moreover, in slightly doped NCCO the volume
of electron pocket (FS patch) is approximately equal to the doping
density x. Our data suggest that the FS patch dominates the
transport in the lightly doped samples around 300 K. It is worthy
to note that the similar situation exists in $La_{2-x}Sr_xCuO_4$
system. In this case, the value of $eR_Hx/V$ is also close to 1
for lightly doping samples \cite{Ando01}, and $eR_Hx/V$ sign
changes in heavily overdoped samples\cite{Hwang}. In ordinary
single band model, the lower limit of $eR_Hx/V$ in hole-doped
sample is $\frac{x}{1-\mid x \mid}$. It should be 0.20, 0.27 and
0.33 for the samples with x=0.17, 0.21 and 0.25, respectively.
Based on the results by Ando et al., it is found that the
experimental lower limit is approximately 0.30, 0.15 and 0.06 for
those cases\cite{note01}. According to above explanation, the
large difference between experimental data and theoretical
prediction for the samples with x=0.21 and 0.25 suggests the
invalidity of "single-band model". One must consider the
competition between electron-like and hole-like FS as the FS
eventually becomes electron-like for overdoped hole-type cuprates
in heavily doped LSCO\cite{Ino,Ando01}.

In conclusion, the Hall angle for the underdoped crystals with
x=0.025 and x=0.06 follows a $T^4$ law, while obeys a $T^2$ law
for the overdoped sample with x=0.2 although the resistivity
remains nearly $T^2$ law in all the samples. This is different
from the hole-doped system. It is closely related to the different
evolution of Fermi surface with doping level. By the delicate
study of $eR_Hx/V$, we also try to understand the behavior of
change of $R_H$ sign in a unified viewpoint for both hole-doped
LSCO and electron-doped cuprates. Our finding confirms that one
must use two-bands model to explain the sign change of $R_H$ and
thermoelectric power with doping. The two-bands model is
associated with the mysterious change of Fermi surface observed by
ARPES.

 This work is supported by the Nature
Science Foundation of China, and by the Knowledge Innovation
Project of Chinese Academy of Sciences and the Ministry of Science
and Technology of China.

\end{document}